\documentclass[twocolumn,superscriptaddress,longbibliography,aps,pra,preprintnumbers,showkeys]{revtex4-1}
\usepackage{hyperref}
\usepackage{epsfig}
\usepackage{graphicx}
\usepackage{epstopdf}
\usepackage{tikz}
\usepackage{float}
\usepackage{amsmath}
\usepackage{amssymb}
\usepackage{bm}
\usepackage[normalem]{ulem}

\begin{document}

\title{Transport properties of proximitized double quantum dots}

\author{G. G\'orski}
\email{ggorski@ur.edu.pl}
\affiliation{Institute of Physics, College of Natural Sciences, University of Rzesz\'{o}w,\\
ul. Pigonia 1, PL-35-310 Rzesz\'{o}w, Poland}

\author{K. Kucab}
\email{kkucab@ur.edu.pl}
\affiliation{Institute of Physics, College of Natural Sciences, University of Rzesz\'{o}w,\\
ul. Pigonia 1, PL-35-310 Rzesz\'{o}w, Poland}

\date{\today}

\begin{abstract}
We study the sub-gap spectrum and the transport properties of a double quantum dot coupled to metallic and superconducting leads. 
The coupling of both quantum dots to the superconducting lead induces a non-local pairing in both quantum dots by the Andreev reflection processes. 
Additionally, we obtain two channels of Cooper pair tunneling into a superconducting lead. In such a system, the direct tunneling process (by one of two dots) or the crossed tunneling process (by both quantum dots at the same time) is possible. We consider the dependence of the Andreev transmittance on an inter-dot tunneling amplitude and the coupling between a quantum dot and the superconducting lead. We also consider the occurrence of interferometric Fano-type line shapes in the linear Andreev conductance spectra. 
  
\end{abstract}


\maketitle

\section{Introduction}
\label{sec:intro}

The rapid development of electronics results in research of the transport properties of different types of heterostructures consisting of nano-objects such as the quantum dots or nanowires. One of the directions of interest is the system consisting of double quantum dots (DQD) placed between superconducting, magnetic or metallic leads \cite{Kim-2001, Cornaglia-2005, Zitko-2006,Tanaka-2008,Sasaki-2009, Zitko-2010, Ferreira-2011, Baranski-2011, Baranski-2012, Tanaka-2012,Siqueira-2012,Calle-2013, Silva-2013,Wojcik-2014, Baranski-2015, Wojcik-2015,Zitko-2015, Wojcik-2016,Calle-2017, Wrzesniewski-2017,Busz-2017, Weymann-2018, Wojcik-2018,Assuncao-2018, Wojcik-2019,Wang-2019,Trocha-2014,Baranski-2020, Shang-2015, Shang-2018}. 

For a system with a DQD coupled to two metallic or ferromagnetic electrodes \cite{Kim-2001, Cornaglia-2005, Zitko-2006,Sasaki-2009,Zitko-2010,Tanaka-2012,Wojcik-2014}, we observe the coexistence of Kondo \cite{Kondo-1964} and Fano \cite{Fano-1961} effects. The connection between one quantum dot ($\mathrm{QD_1}$) and metallic electrodes leads to the widening of a dot level. When the second dot ($\mathrm{QD_2}$) is side coupled to the first one, there is a possibility to obtain the Fano-like asymmetric line shapes in the linear conductance \cite{Zitko-2006,Tanaka-2008,Sasaki-2009, Zitko-2010,Ferreira-2011,Wojcik-2014} which are obtained as a result of the interference between discrete $\mathrm{QD_2}$ level with a broad band of $\mathrm{QD_1}$. Additionally, for interacting dots, one obtains a two-stage Kondo effect \cite{Cornaglia-2005, Zitko-2006,Wojcik-2014, Shang-2015, Shang-2018}. The Fano destructive interference partially suppresses the Kondo resonance.

For the case of one quantum dot attached to one superconducting (SC) and one normal metallic (N) contact (N-QD-SC system), the propagation of a Cooper pair into SC lead and a hole reflection into a metallic lead (Andreev reflection process) occurs in the system \cite{Andreev-1964,Rodero-2011}. The connection of second QD into the N-QD-SC system causes the competition between the Andreev and Fano effect \cite{Tanaka-2008, Baranski-2011, Baranski-2015,Wang-2019,Baranski-2020}. Other options of DQD with SC and N leads connection are also considered, e.g. connection of the first dot to two metallic leads and the second dot to the SC lead \cite{Siqueira-2012,Calle-2013,Calle-2017,Wojcik-2018,Wojcik-2019}. In such systems, there are possible both the normal electron transfers (when the single electron transfers between  both normal metallic electrodes) and direct (DAR) and crossed (CAR) Andreev reflections (when electrons of the Cooper pair tunnel into SC lead and the holes tunnel to the same (DAR) or the second (CAR) metallic lead). The hybrid DQD structure can be used as a Cooper pair splitter if each QD is connected with separate metallic leads \cite{Wrzesniewski-2017,Busz-2017,Assuncao-2018}.    

\begin{figure}[h]
\centering
\includegraphics[width=1\linewidth]{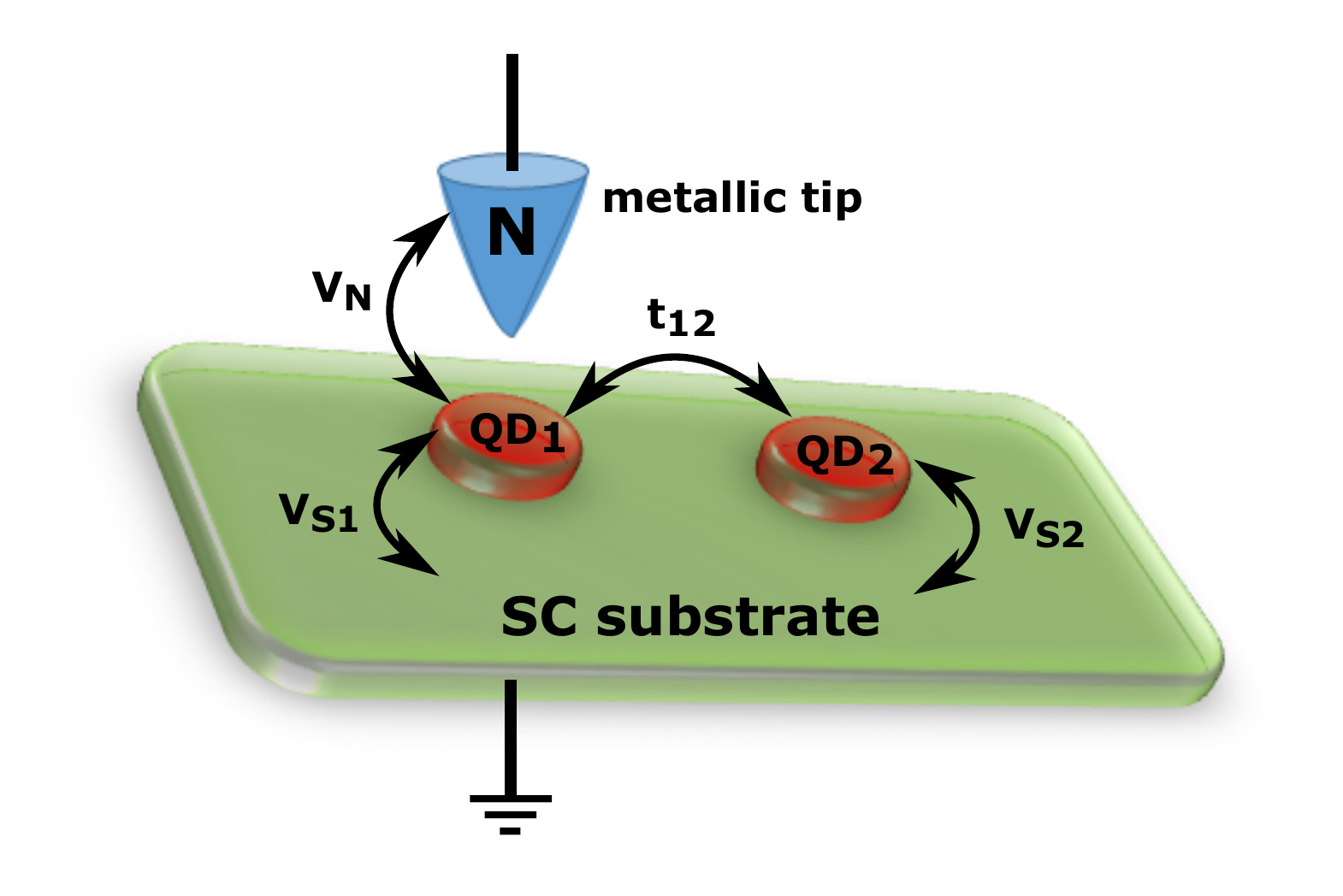}
\caption{Schematic representation of the double quantum dot system. The metallic lead is coupled to the first quantum dot ($\mathrm{QD_1}$), the superconductor substrate
is attached to both quantum dots ($\mathrm{QD_1}$ and $\mathrm{QD_2}$).  }
\label{scheme}
\end{figure}

In this paper we consider a system consisted of two quantum dots embedded in a superconducting substrate (see Fig. \ref{scheme}). 
We assume that a metallic lead is connected with one of these dots. Such a system can be realized experimentally by the use of scanning tunneling microscopy (STM) measurements of metallic atoms (e.g. Fe atoms) embedded in the superconducting substrate (e.g. Pb substrate) by the use of metallic tip.
The STM-based single-atom manipulations technique is currently widely used for the detection of Majorana bound states in metallic chains \cite{Nadj-Perge-2014,Pawlak-2016,Jeon-2017,Ruby-2017,Kim-2018}. This method allows for precise positioning of atoms on the substrate and for the local determination of spectral and transport properties of individual atoms \cite{Kim-2018}. In the system considered by us, the neighborhood of SC substrate with QDs, by the proximity effect, generates the Andreev states both on $\mathrm{QD_1}$ and on $\mathrm{QD_2}$. The coupling of both QDs with SC lead causes that the Cooper pair can tunnel to the SC lead via one of two dots (direct tunneling) or via both dots at the same time (crossed tunneling).
Our aim is to analyze the spectral and transport properties of the quantum dot $\mathrm{QD_1}$ depending on its coupling with proximitized $\mathrm{QD_2}$.
These results are important in the context of the distinction between the coupling of QD with trivial Andreev bound states and topological Majorana bound states \cite{Liu-2018}.

This work is structured as follows. In Sec. \ref{sec:model} we introduce the microscopic model which describes the system considered by us. In this section, we introduce the relations describing the transport properties of the system. In Sec. \ref{sec:results} we present the numerical results for the local density of states and for the Andreev transmission coefficient. We also analyze the influence of our model parameters on the observed Fano-type line shapes of Andreev transmittance.   
Finally, the conclusions can be found in Sec. \ref{sec:conclusions}

\section{The model}
\label{sec:model}

We consider the system consisted of two quantum dots embedded in the superconducting substrate (see Fig. \ref{scheme}). The use of a metallic tip allows us to obtain the current characteristics of the system. We assume that the QDs are connected to the same SC leads, so the difference of the superconducting phases does not occur and the Josephson current is not observed. The total Hamiltonian of our setup has the following form:
\begin{eqnarray}
H&=&\sum_{i\sigma}\varepsilon_i d^{\dagger}_{i\sigma} d_{i\sigma}+\sum_{\sigma}t_{12}(d^{\dagger}_{1\sigma} d_{2\sigma}+h.c.) \\
&+&\sum_{\textbf{k} \sigma \beta}\xi_{\textbf{k} \beta}c^{\dagger}_{\textbf{k} \sigma \beta}c_{\textbf{k}\sigma \beta}-\sum_{\textbf{k}} ( \Delta c^{\dagger}
_{\textbf{k}\uparrow S}c^{\dagger}_{-\textbf{k} \downarrow S}+h.c.) \nonumber \\
&+&\sum_{\textbf{k} \sigma}(V_{\textbf{k} N}d^{\dagger}_{1 \sigma} c_{\textbf{k} \sigma N}+h.c.)
+\sum_{\textbf{k} i \sigma}(V_{\textbf{k} Si}d^{\dagger}_{i \sigma} c_{\textbf{k} \sigma S}+h.c.),\nonumber 
\label{HAnd}
\end{eqnarray}
where $d_{i\sigma}^\dagger(d_{i\sigma})$ are the creation (annihilation) operators of an electron with spin $\sigma$ at $\mathrm{QD_i}$ ($i=1,2$), $\varepsilon_i$ is the energy level of QD$_i$, $t_{12}$ is the inter-dot tunneling amplitude, $c^{\dagger}_{\textbf{k} \sigma \beta}(c_{\textbf{k}\sigma \beta})$ denote the creation (annihilation) operators of an electron with momentum $\textbf{k}$ and spin $\sigma$ in the metallic tip ($\beta=N$) or in the superconducting substrate ($\beta=S$), $\xi_{\textbf{k} \beta}=\epsilon_{\textbf{k} \beta}-\mu_\beta$ is an energy dispersion of the lead $\beta$ measured with respect to the electrochemical potentials $\mu_\beta$. We assume that $\mu_S=0$ and $\mu_N=eV$. $V_{\textbf{k}N}$ is the tunneling amplitude between the QD$_1$ and the metallic tip, and $V_{\textbf{k} Si}$ is the tunneling amplitude between the $i$-dot and superconducting substrate. $\Delta$ is the superconducting energy gap. 

The coupling between QDs and an SC substrate leads to the Andreev reflection processes \cite{Andreev-1964}, where taking an electron from a metallic lead causes the injection of a Cooper pair into the SC lead and the reflection of a hole into a metallic lead. In the considered system, for $V_{kS2} \neq 0$, the direct tunneling is possible when the injection of the Cooper pair occurs from one of two QDs, or crossed tunneling is possible when the Cooper pair creates one electron from each dot.

We focus on the Andreev transport regime, so we use the $\Delta\rightarrow \infty$ limit \cite{Rozhkov-2000,Eldridge-2010,Wrzesniewski-2017}. 
In a wide-bandwidth limit, we introduce the coupling constant between $\mathrm{QD_1}$ and a metallic lead $\Gamma_{N}=2\pi\sum\left|V_{kN}\right|^2\delta\left(\omega-\xi_{kN})\right)$, and the coupling constant between $QD_i$ and superconducting substrate $\Gamma_{Si}=2\pi\sum\left|V_{kSi}\right|^2\delta\left(\omega-\xi_{kS})\right)$.
The effective Hamiltonian takes on the following form:
 \begin{eqnarray}
H_{\rm{eff}}&=&\sum_{i\sigma}\varepsilon_i d^{\dagger}_{i\sigma} d_{i\sigma}
+\sum_{\sigma}t_{12}(d^{\dagger}_{1\sigma} d_{2\sigma}+h.c.)\\
&+&\sum_{\textbf{k} \sigma}\xi_{\textbf{k} N}c^{\dagger}_{\textbf{k} \sigma N}c_{\textbf{k}\sigma N} 
+\sum_{\textbf{k} \sigma}(V_{\textbf{k} N}d^{\dagger}_{1 \sigma} c_{\textbf{k} \sigma N}+h.c.)\nonumber \\
&-& \sum_{i}\frac{\Gamma_{Si}}{2}(d^{\dagger}_{i\uparrow}d^{\dagger}_{i\downarrow} +h.c.)
+\sum_{i}\frac{\Gamma_{Si\bar{i}}}{2}(d^{\dagger}_{i\uparrow}d^{\dagger}_{\bar{i}\downarrow} +h.c.),\nonumber
\label{Heff}
\end{eqnarray}
where $\bar{i}=2$ for $i=1$ and $\bar{i}=1$ for $i=2$, $\Gamma_{Si}$($\Gamma_{Si\bar{i}}$) is the direct (cross) coupling between $i$-dot and the SC substrate. We assume that $\Gamma_{S12}=\Gamma_{S21}=\sqrt{\Gamma_{S1}\Gamma_{S2}}$ \cite{Eldridge-2010,Wrzesniewski-2017,Wojcik-2019}.
  
Using the equation of motion method, we obtain the matrix of Green's functions ${\cal{G}}(\omega)=\langle \langle \Psi ; \Psi^{\dagger} \rangle \rangle$, where $\Psi^{\dagger}=(d_{1\uparrow},d_{1\downarrow}^{\dagger},d_{2\uparrow},d_{2\downarrow}^{\dagger})$, in the following notation  

\begin{widetext}
\begin{eqnarray} 
{\cal{G}}^{-1}(\omega) =
\left( \begin{array}{cccc}  
\omega-\varepsilon_{1}+\frac{i\Gamma_N}{2} & \frac{\Gamma_{S1}}{2} & -t_{12} & \frac{-\Gamma_{S12}}{2}\\
\frac{\Gamma_{S1}}{2}&\omega+\varepsilon_{1}+\frac{i\Gamma_N}{2}&\frac{-\Gamma_{S12}}{2}& t_{12}\\
-t_{12}&\frac{-\Gamma_{S12}}{2}&\omega-\varepsilon_{2}& \frac{\Gamma_{S2}}{2} \\
\frac{-\Gamma_{S12}}{2} &t_{12}&\frac{\Gamma_{S2}}{2} &\omega+\varepsilon_{2} 
\end{array}\right).
\label{Gr44}
\end{eqnarray} 
\end{widetext}

In the SC atomic limit, $\Gamma_{N}\rightarrow 0$, the Green's functions are characterized by four poles 
\begin{eqnarray}
\varepsilon_{A1}=1/\sqrt{2}\sqrt{A + \sqrt{A^2-4B}},\nonumber\\
\varepsilon_{A2}=-1/\sqrt{2}\sqrt{A + \sqrt{A^2-4B}},\nonumber\\
\varepsilon_{A3}=1/\sqrt{2}\sqrt{A - \sqrt{A^2-4B}},\nonumber\\
\varepsilon_{A4}=-1/\sqrt{2}\sqrt{A - \sqrt{A^2-4B}}
\label{EA}
\end{eqnarray}
where $A=\varepsilon_1^2+\varepsilon_2^2+(\frac{\Gamma_{S1}+\Gamma_{S2}}{2})^2+2t_{12}^2$ and $B=(\varepsilon_1\varepsilon_2-t_{12}^2)^2+\left(\frac{\varepsilon_1\Gamma_{S2}+\varepsilon_2\Gamma_{S1}}{2}+\Gamma_{S12}t_{12}\right)^2$. These poles correspond to four Andreev resonances. The non-zero value of $\Gamma_{N}$ causes the broadening of these resonances. The generation of four Andreev states is related to the fact, that the proximity effect generates the Andreev states on both quantum dots. The generation of Andreev states on $\mathrm{QD_1}$ is related to the direct coupling of the $\mathrm{QD_1}$ with the SC lead. For this quantum dot the $\varepsilon_{A1}$ and $\varepsilon_{A2}$ states are dominant. For $\mathrm{QD_2}$ we have two methods of generation of the Andreev states, (i) the direct one for $\Gamma_{S2} \neq 0$; (ii) the indirect one (via $\mathrm{QD_1}$) for $\Gamma_{S2} = 0$ and $t_{12} \neq 0$. For this quantum dot $\varepsilon_{A3}$ and $\varepsilon_{A4}$ states dominate.

The properties of nanoscopic systems can be analyzed experimentally using the current characteristics, especially the zero-bias differential conductance. The current flowing from the N lead can be calculated using the following equation \cite{Jauho-1994,Sun-2000,Yamada-2011}
\begin{eqnarray}
I_N&=&-e \frac{d}{dt}<\sum_{k\sigma}c^\dagger_{k\sigma N}c_{k\sigma N}>\\
&=&-\frac{2e\Gamma_N}{h}\rm{Im}\int{\left[ 2f(\omega-\mu_N){\cal{G}}^r_{11}(\omega)+{\cal{G}}^<_{11}(\omega)\right]d\omega}\nonumber 
\label{Indef}
\end{eqnarray}
where ${\cal{G}}^r ({\cal{G}}^<)$ are the retarded (lesser) Green's functions, respectively, and $f(\omega)$ is the Fermi distribution function. 

The total current is the sum of the normal and Andreev currents. At low temperatures, the normal current, when an electron moves from N lead to SC lead, is realized for $eV\geq \Delta$. For $eV\leq \Delta$ in the N-QD-SC system, we observe the Andreev current arising when an electron from N lead pairs with a second electron with opposite spin and as the Cooper pair they are tunneling to the SC lead and simultaneously, a hole with opposite spin is reflected back to the N lead \cite{Andreev-1964, Deutscher-2005}. 
In our N-DQD-SC system, with $\Gamma_{S1} \neq 0$ and $\Gamma_{S2} \neq 0$, the Cooper pair can tunnel to the SC lead via one of two dots (direct tunneling) or via both dots at the same time (crossed tunneling).

For $\Delta\rightarrow \infty$ in our system there occurs the Andreev current only, so the relation describing the current $I_N$ has the following form \cite{Sun-2000,Baranski-2011}:
 
\begin{eqnarray}
I_N=\frac{e}{h}\int{T_A(\omega)\left[ f(\omega-\mu_N)-f(\omega+\mu_N)\right]d\omega} 
\label{In}
\end{eqnarray}
where 
\begin{eqnarray}
T_A(\omega)=2\Gamma_N^2\left|{\cal{G}}^r_{12}(\omega)\right|^2
\label{TA}
\end{eqnarray}
is the total Andreev transmittance. The maximum of $T_A$ value is equal to 2. The Andreev transmittance is always symmetric, $T_A(\omega)=T_A(-\omega)$, because the anomalous Andreev scattering involves both the particle and hole degrees of freedom. 

\begin{figure*}[th]
\centering
\includegraphics[width=0.9\linewidth]{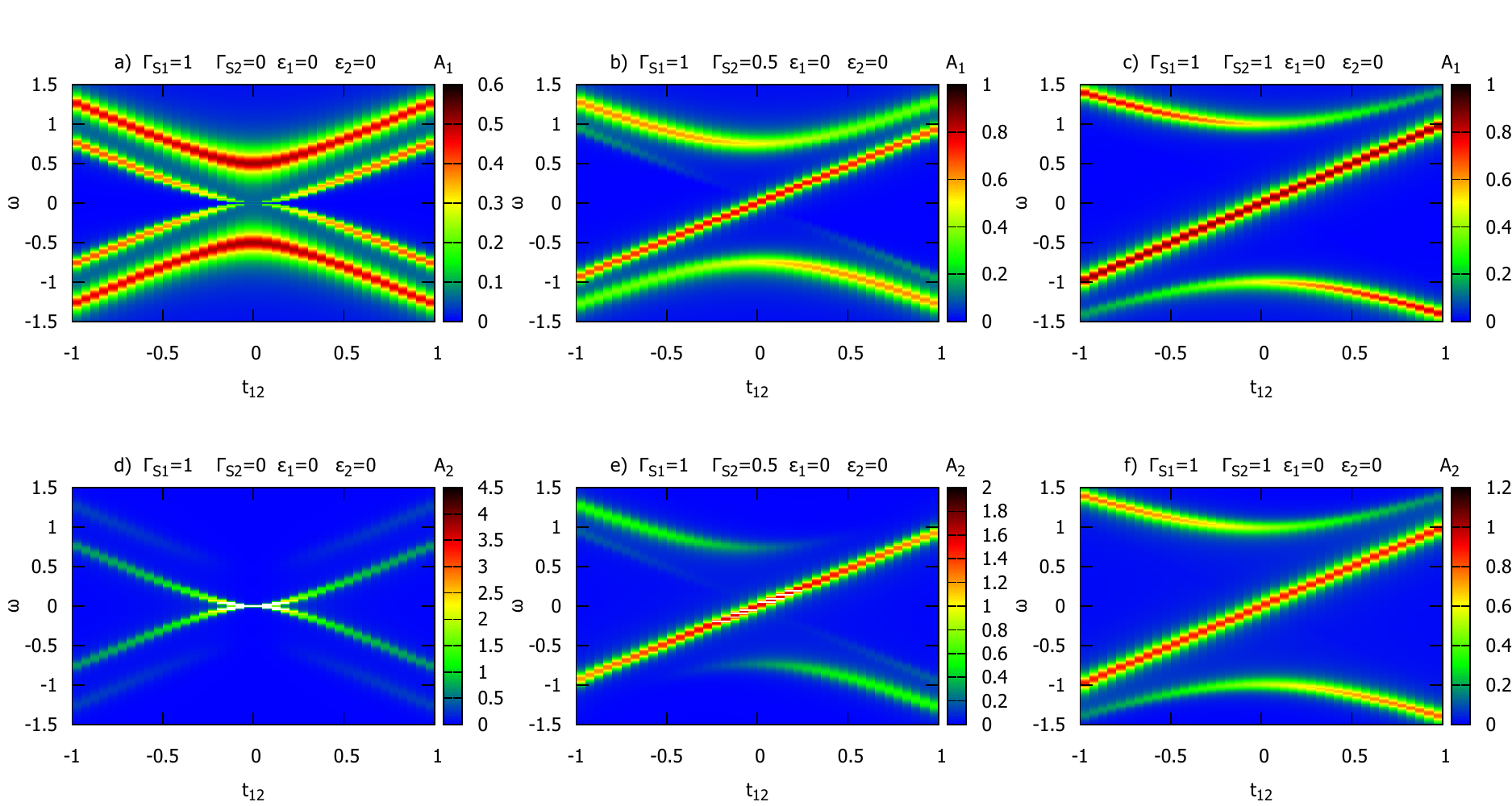}
\caption{The normalized spectral density of quantum dots (for $\mathrm{QD_1}$ - top panel, and for $\mathrm{QD_2}$ - bottom panel) as a function of the inter-dot tunneling amplitude $t_{12}$ and for different values of the coupling parameter $\Gamma_{S2}$. Other parameters are $\varepsilon_{1}=\varepsilon_{2}=0$, $\Gamma_{S1}=1$ and $\Gamma_N=0.25$.}
\label{figrozmt12}
\end{figure*}

The knowledge of an Andreev transmittance allows us to calculate the zero-bias differential conductance as
\begin{eqnarray}
G_A(V=0)&=&\left.\partial I_N/\partial V\right|_{V\rightarrow 0}\\
&=&\frac{2e^2}{h}\int{T_A(\omega)\left[ -\frac{\partial f(\omega)}{\partial \omega}\right]d\omega}.\nonumber 
\label{GA}
\end{eqnarray}

At low temperatures, this equation can be simplified as follows:
\begin{eqnarray}
G_A(V=0)=\frac{2e^2}{h}T_A(0). 
\label{GA0}
\end{eqnarray}

\begin{figure*}[th]
\centering
\includegraphics[width=0.9\linewidth]{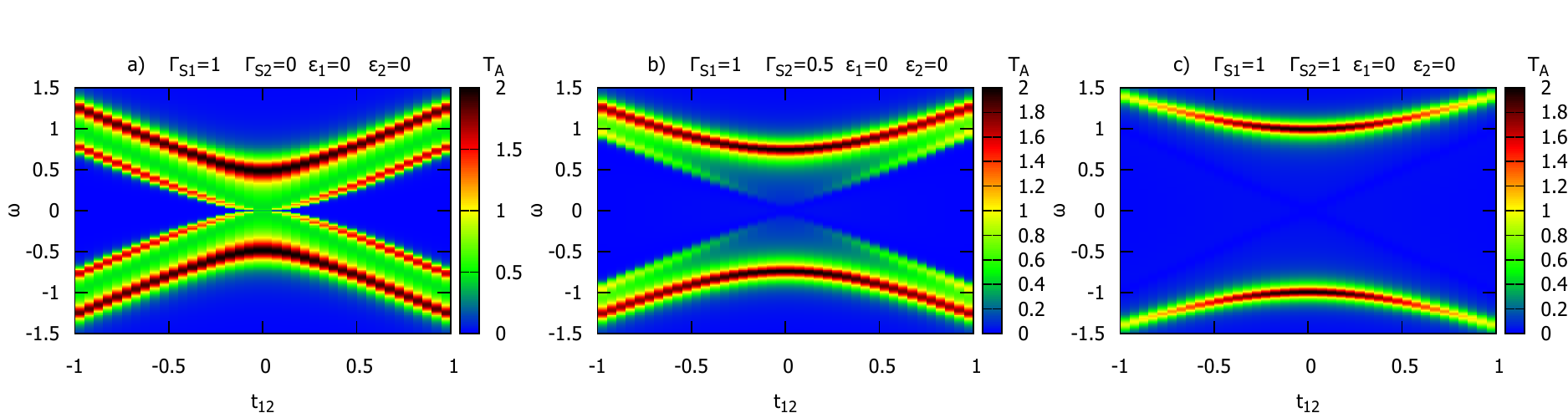}
\caption{The Andreev transmittance as a function of the inter-dot tunneling amplitude $t_{12}$, and for different values of the coupling parameter $\Gamma_{S2}$. Other parameter are $\varepsilon_{1}=\varepsilon_{2}=0$, $\Gamma_{S1}=1$ and $\Gamma_N=0.25$.}
\label{figTAzmt12}
\end{figure*}

\section{Results}
\label{sec:results}

In this section, we present the numerical results for the spectral density of QDs and the Andreev transmittance. As a unit of energy we assume the coupling parameter between $\mathrm{QD_1}$ and SC substrate ($\Gamma_{S1}=1$). The computations were carried out at $T = 0$.

\subsection{Spectral density}
\label{sec:density}

The normalized spectral density of a quantum dot is defined as:
\begin{eqnarray}
A_i(\omega)=-\frac {\Gamma_N}{2} \rm Im <<d_{i\uparrow};d_{i\uparrow}^\dagger>>_\omega.
\label{ro}
\end{eqnarray}

With such defined normalized spectral density, the maximum value of $A_1(\omega)$ is equal to 1.

In Fig. \ref{figrozmt12} we present the normalized spectral density of $\mathrm{QD_1}$ (top panel) and $\mathrm{QD_2}$ (bottom panel) as a function of the inter-dot tunneling amplitude ($t_{12}$) and for different values of the coupling parameter $\Gamma_{S2}$. For the computations, we assumed an equal energy level for QDs ($\varepsilon_1=\varepsilon_2=0$), which is consistent with a chemical level of SC lead. 

For $\Gamma_{S2}=0$ (left panel) we obtain the system with a side-coupled $\mathrm{QD_2}$, which is not directly coupled to the leads (metallic or superconducting) 
\cite{Tanaka-2008, Baranski-2011, Tanaka-2012,Silva-2013,Baranski-2015, Wojcik-2015}. In this case, the dependence of spectral density for $\mathrm{QD_1}$ and $\mathrm{QD_2}$ is symmetric ($A_i(\omega)=A_i(-\omega)$). We also observe the symmetry of spectral density as a function of the inter-dot tunneling amplitude with respect to $t_{12}=0$.
As Bara\'{n}ski and Doma\'{n}ski shown \cite{Baranski-2011}, in this T-shape configuration the Fano-type resonances and antiresonances localized near $\pm \varepsilon_{2}$ are obtained. The Fano resonances are characterized by a typical asymmetric line shape and are obtained if a broad spectrum interferes with a discrete level.
In our system, a broad $\mathrm{QD_1}$ spectrum, resulting from coupling $\mathrm{QD_1}$ with metallic lead, interferes with discrete $\mathrm{QD_2}$ level. As a result of Fano type quantum interference, for $\omega=\varepsilon_2$ we obtain the spectral density $A_1(\omega=\varepsilon_2)=0$ for all values of $t_{12} \neq 0$. The location of this antiresonance also does not depend on $\varepsilon_{1}$. At strong inter-dot tunneling ($t_{12}>\Gamma_N$), both for $A_1(\omega)$ and $A_2(\omega)$, one can observe four resonance states localized near energies $\varepsilon_{Ai}$ (see Eq. \ref{EA}), which come from a direct Andreev effect for $\mathrm{QD_1}$ and from an indirect Andreev effect for $\mathrm{QD_2}$ \cite{Baranski-2011}. For $\varepsilon_1=\varepsilon_2=0$ two outer states (localized near $\varepsilon_{A1}$ and $\varepsilon_{A2}$) dominate for $A_1(\omega)$, while two inner states (localized near $\varepsilon_{A3}$ and $\varepsilon_{A4}$) dominate for $A_2(\omega)$.

\begin{figure*}[thb]
\centering
\includegraphics[width=0.9\linewidth]{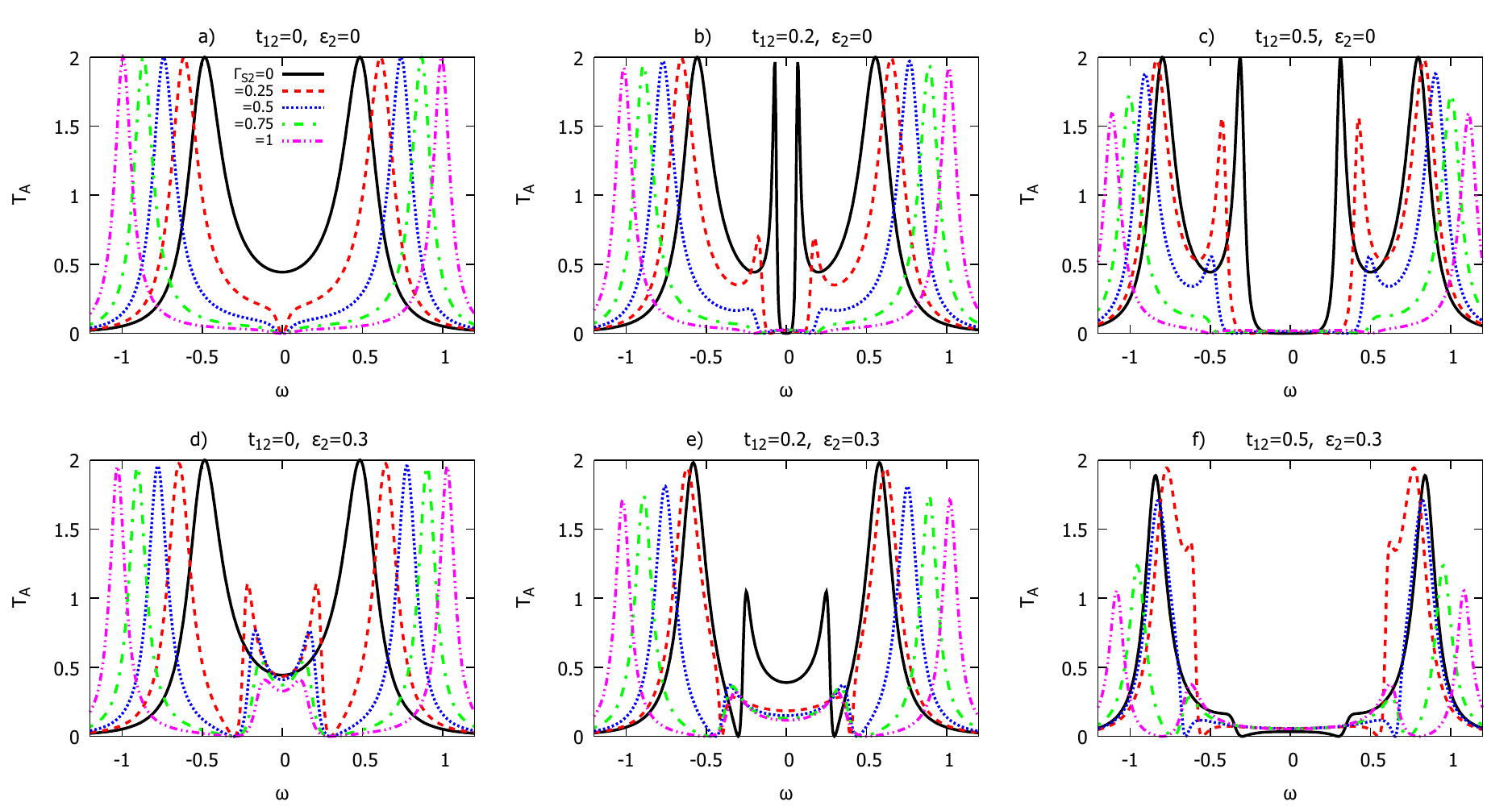}
\caption{The Andreev transmittance  as a function of the coupling parameter $\Gamma_{S2}$ and for different values of the inter-dot tunneling amplitude $t_{12}$ and $\varepsilon_{2}$. 
Other parameter are $\varepsilon_{1}=0$, $\Gamma_{S1}=1$ and $\Gamma_N=0.25$.}
\label{figTAzmGS2}
\end{figure*}

The finite value of coupling between $\mathrm{QD_2}$ and SC lead (middle and right panel of Fig. \ref {figrozmt12}) causes a direct induction of pairing on $\mathrm{QD_2}$. 
In this case, despite the dots energies which are equal to 0, the symmetry breaking of $A_i(\omega)$ is observed, whereas one can observe the $A_i(\omega,t_{12})=A_i(-\omega,-t_{12})$ dependence. The outer Andreev states are still highly visible, and their location depends both on $\Gamma_{S1}$ and $\Gamma_{S2}$. 
For the inner states, we obtain a strong peak near $\varepsilon_{A3}$ and very weak peak near $\varepsilon_{A4}$. The non-zero value of $\Gamma_{S2}$ causes that 
we do not observe the Fano-type line shape of $A_1(\omega)$ ($A_1(\omega=\varepsilon_2) \neq 0$). For identical quantum dots with $\varepsilon_1=\varepsilon_2$, which are characterized by an identical coupling with SC substrate $\Gamma_{S1}=\Gamma_{S2}$, we obtain three-center structure with one pair of Andreev resonances 
localized near $\varepsilon_{A1}$ and $\varepsilon_{A2}$, and with strong resonance near $\varepsilon_2+t_{12}$ (see right panel of Fig. \ref {figrozmt12}).  

\subsection{Andreev transmittance}
\label{sec:transmittance}

The dependence of Andreev transmittance (Eq. \ref{TA}) as a function of inter-dot tunneling amplitude $t_{12}$ is shown in Fig. \ref{figTAzmt12}. 
In the case of side-coupled $\mathrm{QD_2}$ ($\Gamma_{S2}=0$ and $t_{12} \neq 0$), two pairs of resonance states are visible in Andreev transmittance (see Fig. \ref{figTAzmt12}(a)) near $\omega=\varepsilon_{Ai}$. The broad resonances are obtained for $\omega=\varepsilon_{A1}$ and $\omega=\varepsilon_{A2}$ and the narrow resonances are obtained for $\omega=\varepsilon_{A3}$ and $\omega=\varepsilon_{A4}$. Taking into account the coupling between $\mathrm{QD_2}$ and a superconducting lead ($\Gamma_{S2} \neq 0$), one obtains the extinction of inner Andreev transmittance resonances (see Figs \ref{figTAzmt12}(b) and (c)).
The increase of $\Gamma_{S2}$ causes the shift of Andreev transmittance resonances localized near $\varepsilon_{A1}$ and $\varepsilon_{A2}$ towards higher energies. Additionally, these resonances become narrower. For the identical coupling of quantum dots with SC substrate, $\Gamma_{S1}=\Gamma_{S2}$, we obtain the total extinction of inner Andreev transmittance resonances (see Fig \ref{figTAzmt12} (c)).

Now we will discuss the influence of the hybridization parameter $\Gamma_{S2}$ on the Andreev transmittance (Fig. \ref{figTAzmGS2}). In our analysis we consider the system of two QDs with equal energy $\varepsilon_1=\varepsilon_2=0$ (top panel) and with different energies $\varepsilon_1=0$ and $\varepsilon_2=0.3$ (bottom panel).
For $\Gamma_{S2}=0$ (solid black line) we obtain the double QDs coupled in a T-shape configuration with metallic and superconducting lead \cite{Tanaka-2008, Baranski-2011, Baranski-2015}. As Bara\'{n}ski and Doma\'{n}ski \cite{Baranski-2011} shown, in this configuration, for small $t_{12}\ll \Gamma_N$ one can obtain the  Fano-type line shapes of Andreev transmittance. At high values of $t_{12}$, the Fano-type features disappear, evolving into the new quasi-particle peaks. Additional peaks can be interpreted as the Andreev peaks being a consequence of the indirect proximity effect induced by $\mathrm{QD_1}$ on the side-attached $\mathrm{QD_2}$. In this configuration, generally, for all values of $t_{12} \neq 0$, one obtains the $T_A(\varepsilon_2)=T_A(-\varepsilon_2)=0$ dependence. 

For a double quantum dot system, the Fano resonance is molded by the coupling of a narrow level related to $\mathrm{QD_2}$, and a broad level related to $\mathrm{QD_1}$. 
In the case of direct coupling of $\mathrm{QD_2}$ with SC lead $\Gamma_{S2} \neq 0$, as a result of proximity effect, there are created the narrow resonance Andreev levels on $\mathrm{QD_2}$. In this case, we obtain the zero-value of Andreev transmittance (Fano dip) for energy value equal to

\begin{eqnarray} 
\varepsilon_F=\pm\sqrt{\varepsilon_2^2+\frac{2t_{12}\Gamma_{S12}\varepsilon_2+\Gamma_{S2}t_{12}^2}{\Gamma_{S1}}}.
\label{EF}
\end{eqnarray}

As we have shown in Fig. \ref{figTAzmGS2} (a) and (d), the zeroing of Andreev transmittance ($T_A(-\varepsilon_F)=T_A(\varepsilon_F)=0$) does not require $t_{12} \neq 0$. At $t_{12}=0$ and $\Gamma_{S2} \neq 0$, the zero value of Andreev transmittance is obtained for $\varepsilon_F=\pm\varepsilon_2$, while for $t_{12} \neq 0$ we obtain that $\varepsilon_F$ is not constant for a given $\varepsilon_2$ but it also depends on $\Gamma_{S1}$, $\Gamma_{S2}$, $\Gamma_{S12}$ and $t_{12}$ (see Fig. \ref{figTAzmGS2} (b), (c), (e) and (f)).   

\begin{figure*}[th]
\centering
\includegraphics[width=0.85\linewidth]{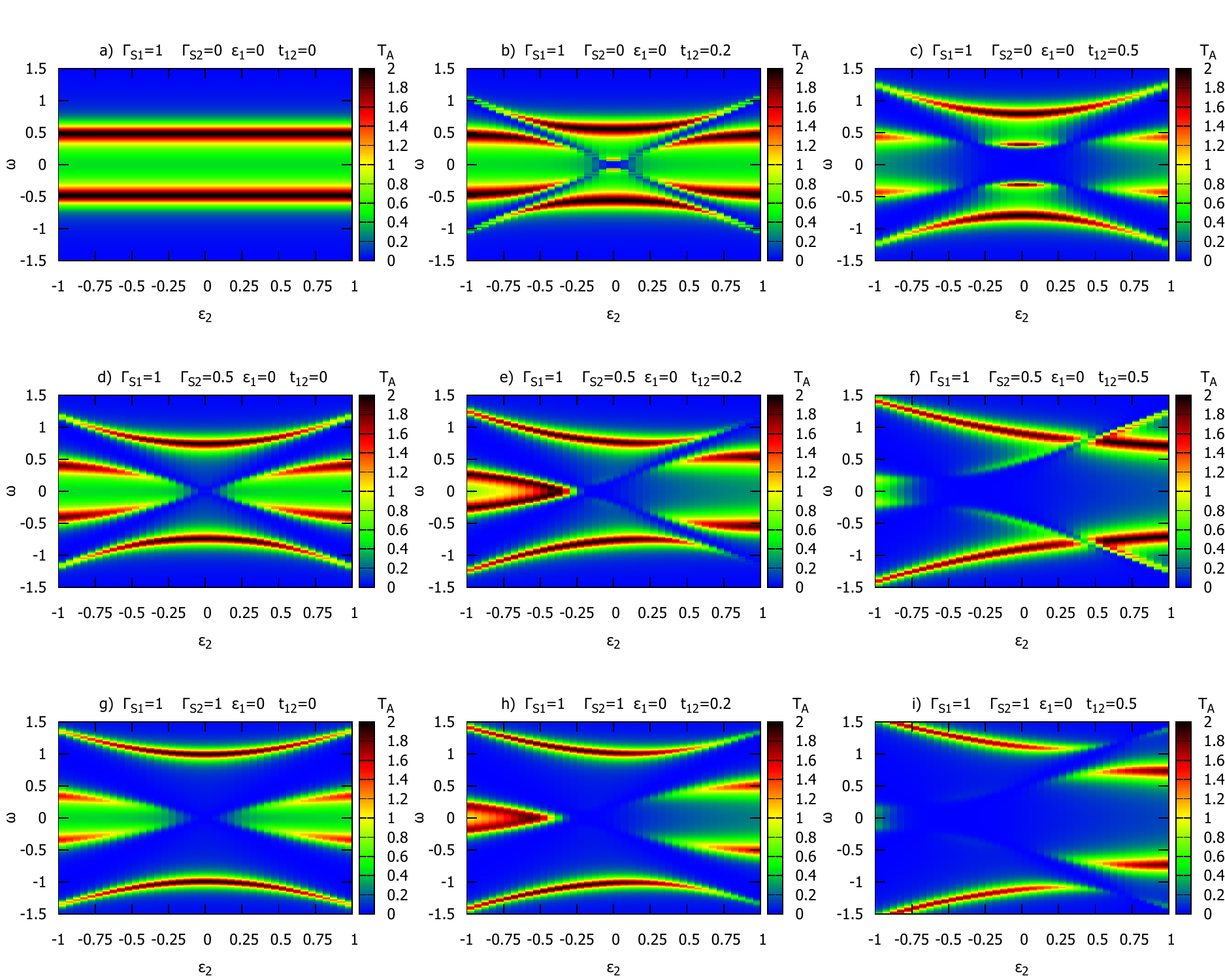}
\caption{The Andreev transmittance as a function of the $\mathrm{QD_2}$ energy and for different values of the inter-dot tunneling amplitude $t_{12}$ and coupling parameter $\Gamma_{S2}$. 
Other parameters are $\varepsilon_{1}=0$, $\Gamma_{S1}=1$ and $\Gamma_N=0.25$.}
\label{figTAzmed2}
\end{figure*}

For QDs which do not interact directly ($t_{12}=0$), the increase of $\Gamma_{S2}$ causes the broadening of an effective value of the coupling parameter $\Gamma_{S\rm{eff}}=\Gamma_{S1}+\Gamma_{S2}$, and as the effect, it causes the shift of Andreev resonances localized near $\omega=\varepsilon_{A1}$ and $\omega=\varepsilon_{A2}$, towards higher energy levels (see Fig. \ref{figTAzmGS2} (a) and (d)). In this case, the maximum of Andreev transmittance is close to 2. For $t_{12} \neq 0$ the increase of $\Gamma_{S2}$ causes the decreasing of the maximum value of Andreev transmittance ($T_A(\varepsilon_{A1})<2$ and $T_A(\varepsilon_{A2})<2$).

In Fig. \ref{figTAzmed2} we present the dependence of Andreev transmittance as a function of $\mathrm{QD_2}$ energy for different values of inter-dot tunneling amplitude $t_{12}$ and for coupling parameter $\Gamma_{S2}$. For $t_{12}=0$ and $\Gamma_{S2}=0$ (Fig. \ref{figTAzmed2} (a)) the transmittance is independent of $\varepsilon_2$.
The non-zero value of $t_{12}$ or $\Gamma_{S2}$ (Figs \ref{figTAzmed2} (b)-(i)) causes that $T_A$ is $\varepsilon_2$ dependent. For almost all values of $\varepsilon_2$ we obtain four resonances of Andreev transmittance.  

For $t_{12} \neq 0$ and $\Gamma_{S2}=0$ (i.e. for a T-shape configuration \cite{Baranski-2011,Tanaka-2008}) or for $t_{12}=0$ and $\Gamma_{S2} \neq 0$ (i.e. without direct coupling between QDs) we obtain $T_A(\omega,\varepsilon_2)=T_A(\omega,-\varepsilon_2)$ (see Figs \ref{figTAzmed2} (b)-(c) and Figs \ref{figTAzmed2} (d),(g), respectively). In the case of $t_{12} \neq 0$ and $\Gamma_{S2}=0$ the zero value of Andreev transmittance is obtained for $\omega=\pm \varepsilon_2$ 
($T_A(\varepsilon_2)=T_A(-\varepsilon_2)=0$). One can see that the clear four resonances of Andreev transmittance are visible for any value of $\varepsilon_2$ (Figs \ref{figTAzmed2} (b)-(c)). For $t_{12} \neq 0$ and $\Gamma_{S2} \neq 0$ (Figs \ref{figTAzmed2} (e)-(f),(h)-(i)) the dependence $T_A(\omega,\varepsilon_2)$ is not symmetrical with respect to $\varepsilon_2=0$. For small values of $t_{12}$, the inner Andreev resonances come close to each other and, in effect, one obtains with properly chosen $\varepsilon_2<0$, one very strong peak of transmittance, close to 2, for $\omega=0$. 

\begin{figure}[h]
\centering
\includegraphics[width=0.85\linewidth]{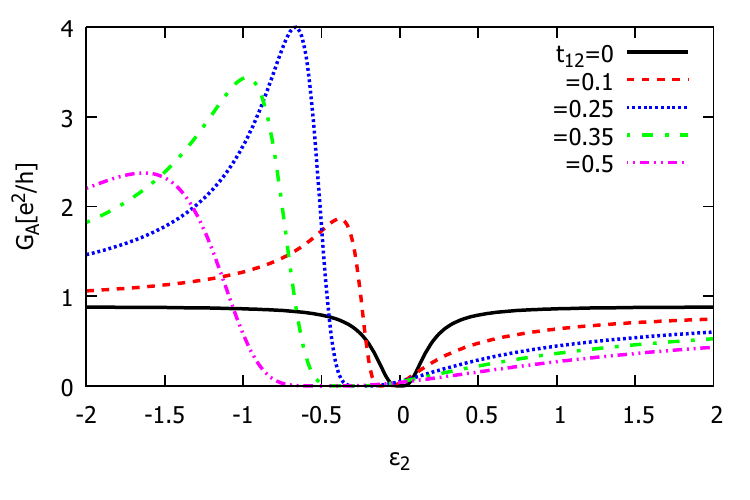}
\caption{The zero-bias Andreev conductance as a function of the $\mathrm{QD_2}$ energy and for different values of the inter-dot tunneling amplitude $t_{12}$. 
Other parameters are $\varepsilon_{1}=0$, $\Gamma_{S1}=\Gamma_{S2}=1$ and $\Gamma_N=0.25$.}
\label{figGAzmed2}
\end{figure}

The formation of a strong peak for $T_A(\omega=0,\varepsilon_2)$ is of great importance for the zero-bias Andreev conductance $G_A=\left.\partial I_N/\partial V\right|_{V\rightarrow 0}$. In Fig. \ref{figGAzmed2} we show the dependence of $G_A$ as a function of $\mathrm{QD_2}$ energy for different values of inter-dot tunneling amplitude $t_{12}$. We have used the symmetric coupling of QDs with SC lead, $\Gamma_{S1}=\Gamma_{S2}=1$. For $\varepsilon_2<0$ and properly chosen value of $t_{12}$, we obtain the maximum value of $G_A=4 e^2/h$. For $\varepsilon_2=-t_{12}$ we obtain $G_A=0$. 

These results can be compared to the zero-bias Andreev conductance results obtained for the N-QD-SC system which is coupled with the topological nanowire hosting Majorana modes (Majorana nanowire) \cite{Baranski-2017,Gorski-2018}. In such a system, for long Majorana nanowire, the optimal value of zero-bias conductance $G_A$ is equal to $1/4$ of the quantum dot's conductance without the coupling with Majorana wire, $\left.G_A\right|_{t_M\neq0}=\frac{1}{4}\left.G_A\right|_{t_M=0}$\cite{Gorski-2018}. In the case of $\mathrm{QD_1-QD_2}$ coupling, there occurs the total reduction of conductance, $G_A=0$. It can be stated that only a fractional reduction of the zero-bias conductance testifies to the fractional fermion character of the Majorana mode. The measurement of the zero-bias Andreev conductance can be treated as the method which allows us to distinguish between the coupling of a QD with the Majorana nanowire (which is characterized by the fractional reduction of $G_A$) and the coupling of a QD with the second QD (where $G_A=0$).

\begin{figure}[h]
\centering
\includegraphics[width=0.8\linewidth]{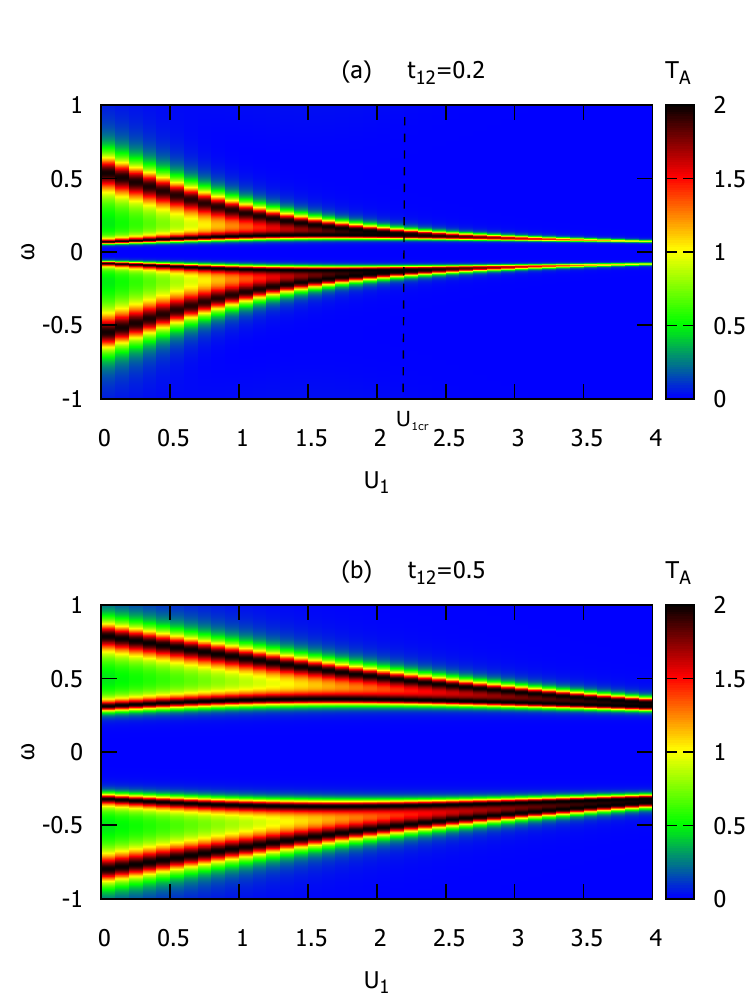}
\caption{The Andreev transmittance as a function of energy $\omega$ and the $\mathrm{QD_1}$ Coulomb interaction $U_1$ for different values of the inter-dot tunneling amplitude $t_{12}$. Other parameters are $\varepsilon_{1}=-U_1/2$, $\varepsilon_{2}=0$ $\Gamma_{S1}=1$, $\Gamma_{S2}=0$ and $\Gamma_N=0.25$. The black dashed line marks $U_{1cr}$ - see the text.}
\label{figTAzmU1}
\end{figure}

\begin{figure}[h]
\centering
\includegraphics[width=0.8\linewidth]{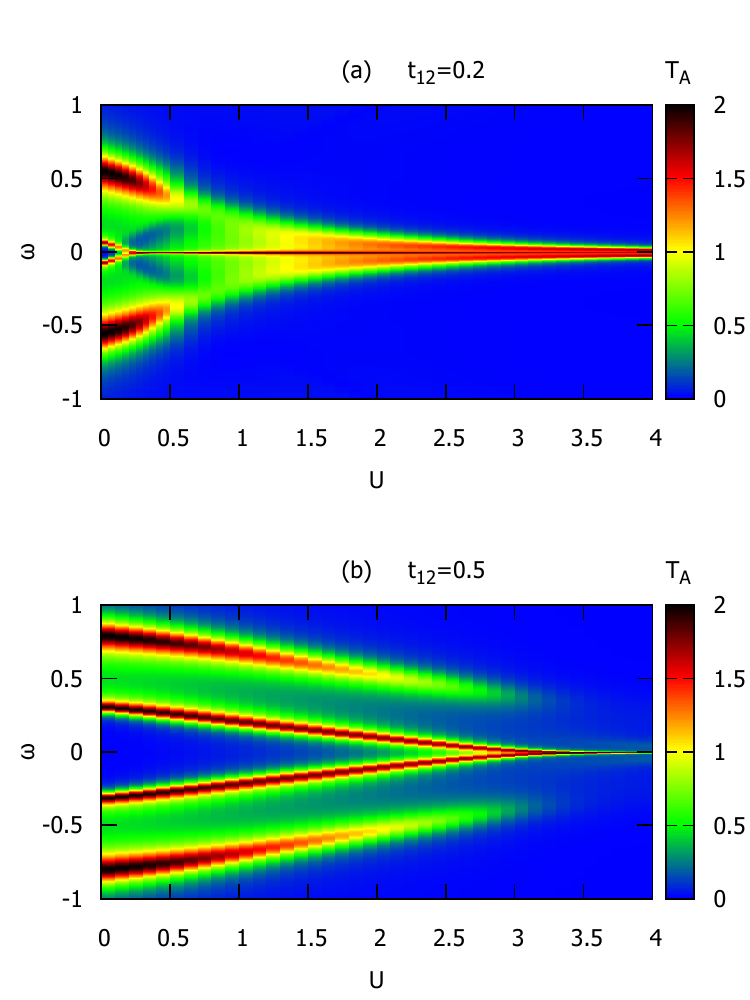}
\caption{The Andreev transmittance as a function of energy $\omega$ and the Coulomb interaction $U_1=U_2=U$ for different values of the inter-dot tunneling amplitude $t_{12}$. 
Other parameters are $\varepsilon_{1}=-U/2$, $\varepsilon_{2}=-U/2$ $\Gamma_{S1}=1$, $\Gamma_{S2}=0$ and $\Gamma_N=0.25$.}
\label{figTAzmU}
\end{figure}

\subsection{The influence of Coulomb interaction}

The Coulomb interaction is very important, taking into account the spectral and transport properties of a quantum dot connected to the SC and metallic leads \cite{Tanaka-2007, Deacon-2010, Rodero-2011, Yamada-2011,Zitko-2015B,Domanski-2016, Hao-2011}. For the systems with a weak coupling of QD with metallic lead, the increase of Coulomb interaction causes the quantum phase transition between the (spin-less) BCS-like singlet and the (spin-full) doublet configurations. This transition takes place for the values of Coulomb interaction close to the QD-SC coupling constant, $\Gamma_S$. For $U<\Gamma_S$ one observes the pair of Andreev peaks in the spectral function. The increase of $U$ interaction causes the split of Andreev peaks, which give rise to the quasi-particle branches $\pm U/2 \pm E_d$, where $E_d = \sqrt{(\varepsilon+U/2)^2+\Gamma_S^2/4}$. The inner low energy peaks approach each other, and for $U>\Gamma_S$ we obtain the strong, central Kondo peak for $\omega=0$ and two Hubbard peaks near $\varepsilon$ and $\varepsilon+U$. 

The Coulomb interaction also modifies the Andreev transmittance of the N-QD-SC system. For $U<\Gamma_S$ we obtain two transmittance peaks. The increase of $U$ values close to $\Gamma_S$ causes the connection of these peaks to one central peak. For $U>\Gamma_S$ the disappearance of transmittance's central peak occurs.   

Now, we will analyze the influence of the Coulomb interaction on the transport properties of a N-DQD-SC system using the second-order perturbation theory \cite{Gorski-2018}. In our analysis we will consider two cases: 
(i) the Coulomb interaction exists only for $\mathrm{QD_1}$ electrons ($U_1 \neq 0$ and $U_2=0$); 
(ii) the Coulomb interaction exists for both quantum dots ($U_1=U_2=U \neq 0$). 
Additionally, we will focus on the particle-hole symmetry case, i.e. when $\varepsilon_1=-U_1/2$ and $\varepsilon_2=-U_2/2$.
In the first case, we will assume that the $\mathrm{N-QD_1-SC}$ system, consisted of correlated $\mathrm{QD_1}$ with $U_1 \neq 0$, is additionally connected to the uncorrelated $\mathrm{QD_2}$.
In fig. \ref{figTAzmU1} we show the dependence of $T_A(\omega)$ as a function of $U_1$ for different values of the inter-dot tunneling amplitude $t_{12}$.
For all values of $U_1$ we obtain $T_A(0)=0$. This result shows that a competition between the Fano and Kondo effects does not allow for the formation of the central Kondo peak for large values of $U_1$ interaction. The increase of $U_1$ interaction causes that the Andreev peaks $\varepsilon_{A1}$ and $\varepsilon_{A2}$ are getting closer but they will never connect each other. The location of $\varepsilon_{A3}$ and $\varepsilon_{A4}$ Andreev peaks, related to the proximity effect on $\mathrm{QD_2}$, slightly shifts when the $U_1$ interaction increases. With properly selected value of $U_1=U_{1cr}$ interaction, the overlap of $\varepsilon_{A1}$ and $\varepsilon_{A3}$, and also $\varepsilon_{A2}$ and $\varepsilon_{A4}$ peaks occur, and as a result, for $U_1>U_{1cr}$, we observe two narrowing peaks. The value of $U_{1cr}$ increases as the $t_{12}$ increases. In this case, the Fano-type resonance plays the dominant role.

Now, we will show the influence of the Coulomb interaction on the Andreev transmittance in a $U_1=U_2=U$ case (see Fig. \ref{figTAzmU}). In this case, the increase of the Coulomb interaction causes that $T_A(0) \neq 0$ for all values of $U \neq 0$. Additionally, one can see that the Andreev peaks are getting closer to each other. For large values of $t_{12}$ and $U$ interactions, the Andreev peaks localized near $\varepsilon_{A1}$ and $\varepsilon_{A2}$ disappear, while the peaks localized near $\varepsilon_{A3}$ and $\varepsilon_{A4}$ are still visible.

\section{Conclusions}
\label{sec:conclusions}

We have analyzed the spectral density and the transport properties of a double quantum dot. Both dots were coupled to a superconducting lead, while only one of them ($\mathrm{QD_1}$) was coupled to the metallic lead. The coupling of both quantum dots with SC lead allows for direct transport of the Cooper pair to the SC lead through one of two quantum dots and for a crossed transport via both quantum dots simultaneously.

The shape of spectral densities of quantum dots strongly depends on the coupling between $\mathrm{QD_2}$ and SC lead. For $\Gamma_{S2}=0$, the spectral densities of quantum dots show the existence of two pairs of Andreev resonances. Additionally, we obtain the Fano dip near $\omega=\varepsilon_2$. 

The non-zero coupling of $\mathrm{QD_2}$ with SC lead causes the extinction of the inner Andreev resonances. Additionally, $\Gamma_{S2} \neq 0$ causes the vanishing of the Fano dip. In the case of $\Gamma_{S2}=\Gamma_{S1}$ and $\varepsilon_1=\varepsilon_2$, we obtain a three-center structure of spectral density consisting of one pair of Andreev resonances and a strong resonance peak near $\varepsilon_2+t_{12}$.
 
For $\Gamma_{S2}=0$, the Andreev transmittance shows one pair of broad peaks localized near $\varepsilon_{A1}$ and $\varepsilon_{A2}$, and narrow resonances near 
$\varepsilon_{A3}$ and $\varepsilon_{A4}$. In this case, the Andreev transmittance shows the Fano dip near $\pm \varepsilon_2$.  

The non-zero value of $\Gamma_{S2}$ causes the extinction of the inner Andreev transmittance resonances. In this case, the Fano dip are still visible, but its location depends on $\varepsilon_2$, $\Gamma_{S1}$, $\Gamma_{S2}$ and $t_{12}$. The level of $\mathrm{QD_2}$ and $t_{12}$ allows for strong modification of the zero-bias Andreev conductance value, $G_A$. For $\varepsilon_2+t_{12}=0$ we obtain the total reduction of $G_A$, while  for properly selected values of $\varepsilon_2$ and $t_{12}$ we can obtain very high values of $G_A$ near $4 e^2/h$.

\begin{acknowledgments}
This project has been supported by the Center for Innovation and Transfer of Natural Sciences and Engineering Knowledge of Rzesz\'{o}w University.
\end{acknowledgments}

\bibliography{myBib}
\end{document}